\documentclass[smallextended,formal]{svjour3}
\usepackage{graphicx}
\begin{document}
\title{Geometric phase as a determinant  of a  qubit--environment coupling}

\author{J. Dajka \and J. {\L}uczka \and P. H\"{a}nggi}

\institute{J. Dajka  \and J. {\L}uczka \at Institute of Physics, University of Silesia, 40-007 Katowice, Poland \and P. H\"{a}nggi \at  Institute of Physics, University of Augsburg,
D-86135 Augsburg, Germany}
\maketitle
\begin{abstract}{
We  investigate   the   qubit geometric phase  and its properties  in dependence on 
the  mechanism for decoherence of a qubit  weakly coupled to its environment.  We consider  two sources of decoherence:  dephasing coupling (without  exchange of energy with environment)  and  dissipative
coupling  (with exchange of energy).  
Reduced dynamics  of  the qubit  is studied in terms of the rigorous  Davies Markovian quantum master equation, both at zero and non--zero temperature. 
 For pure dephasing
coupling,  the geometric phase varies monotonically  with respect to the polar angle (in the  Bloch 
sphere  representation) parameterizing  an  initial state of the qubit. Moreover, it is antisymmetric about some points on the geometric phase-polar angle plane.     This is in
distinct contrast to the case of   dissipative coupling for which the
variation of the geometric phase with respect to the polar angle typically is  non-monotonic, displaying local extrema and is not antisymmetric.
  Sensitivity  of  the geometric phase  to  details of  the decoherence source  can make  it  
 a tool  for testing  the nature of the qubit--environment interaction. }
\end{abstract} 


\section{Introduction}
One of the key obstructions of  an effective implementation of
quantum algorithms is related to the ubiquitous  problem of
decoherence in real quantum objects \cite{nielsen}.  Quantum
decoherence is generic as it results from the imperfect isolation of
the quantum system from its environment. Decoherence can be
diminished under very special conditions such as e.g. the presence
of the decoherence free subspaces \cite{decohfree} or via the
application of tailored, external control schemes \cite{kohler}. A
promising novel direction in quantum information  relates to  so
called holonomic or topological quantum computations
\cite{zanardi,topol} allowing for a substantial reduction of
decoherence \cite{hqc,aqc}. The essence of this method consists in
encoding  the information in the holonomy related to the geometric
phase of the quantum evolution  \cite{berry}. 
The geometric phase can be expressed as a path integral and via
the Stokes theorem, can be  converted into a surface integral.
Therefore, it behaves like a geometric area. A quantity like an area
is  less dependent on the details of time evolution and therefore is
less affected by changes of environmental conditions or an imperfect
control, and hence, is typically more robust. This is the key
attribute that makes  geometric phases attractive  for the
implementations of fault-tolerant quantum computation.  Some
suggestions  have been presented to realize this objective, e.g. in
NMR experiments \cite{hqc}, ion traps \cite{duan}, neutral atoms in
cavity QED \cite{recati}, quantum dots \cite{yellow} or  Josephson
junction devices \cite{falci}. The performance of holonomic quantum
gates under various conditions has been studied recently
\cite{parodi}. 

The  quantum evolution in the presence of decoherence is generically
non--unitary. Therefore,  the notion of geometric phase needs to be
extended.   There are several extensions of the
geometric phase concept for systems which are either in a mixed
state or/and undergo a non--unitary evolution. The first attempt towards
this goal is given in \cite{armin}, being rather of  mathematical
character.  The other are   based on
quantum trajectories \cite{traj}, quantum interferometry \cite{sjuk1} and the state
purification (kinematic approach) \cite{muki,sjuk2}. For non--unitary quantum evolution there
is no commonly accepted scheme of defining the geometric phase in
open quantum systems  \cite{nowe}. 
Here we use the approach based on state purification 
as proposed in Ref. \cite{sjuk2}. This so defined geometric phase
has been extensively studied in various contexts \cite{ph1,ph2}.
 One of the appealing 'advantages' of studying
the phase defined in \cite{sjuk2} is that it can be measured with  a
carefully prepared interferometric experiment \cite{sjuk1,sjuk2}.
Our reasoning is thus guided by its potential for experimental
implementation. 

There is no unique method of describing the time evolution of open
quantum systems and there are  several  schemes to treat such
systems which however typically give rise to non-equivalent dynamics
\cite{chaos,ali1}. One scheme  consists in the  derivation of a
reduced system dynamics, via tracing over the degrees of freedom of
the environment. Except some few exactly solvable models
\cite{chaos,ali} it is not clear how to relate  the reduced dynamics
to the microscopically  first principles dynamics based upon the
Hamiltonian structure of quantum dynamics \cite{chaos,romero}. The
exactly solvable   model of pure dephasing has  been applied for
studying quantum channels \cite{anizfid} or in exploring the
dynamics of quantum entanglement \cite{abel}.   Despite its
simplicity the highly non--trivial properties of geometric phase of
qubits has also been discussed \cite{myfaz}. One of the most
successful examples of constructing reduced dynamics is the Davies
approximation scheme \cite{davies}. Within this approach,
starting with the general 'system--bath--interaction' Hamiltonian,
one obtains, in a mathematically rigorous way, a Markovian master
equation form of a quantum system weakly coupled to the environment
which preserves positivity and yields the correct equilibrium Gibbs
state \cite{spohn}. This approach has been applied to various
problems of statistical physics, quantum optics, solid state physics
and quantum information, e. g. for studying entanglement dynamics in
bipartite systems \cite{lendi}.

In this paper we apply the Davies master equation  to study the
geometric phase of the qubit coupled to the bosonic bath.  Various
families of a coupling and different coupling strength are shown to
result in a qualitative and quantitative modification of the
geometric phase. This behavior could suggest a method to
resolve  the nature of the qubit-bath coupling: In particular, the
dephasing coupling presents not only a mere theoretical construction
but can be realized in experiments within tailored regimes
\cite{devoret}. 
In order to keep this study  self--contained, we briefly review the
notion of the geometric phase for a non--unitarily evolving qubit
and then  present the qubit master equation derived from the
Davies theory.


\section{Geometric phase}

Generally, the time evolution of the qubit reduced density matrix
$\rho(t)$ is neither unitary nor Markovian \cite{chaos,romero}. It
is constructed as the mapping
\begin{eqnarray}\label{general}
\rho(t)=\Lambda(t,t_0)\rho(t_0)
\end{eqnarray}
obeying some properties depending on the specific circumstances and
approximations such as e.g.   the celebrated complete positivity
condition \cite{alickilendi}.  In order to exploit the approach  to the geometric phase based on
state purification \cite{sjuk2} we have to present the reduced
density matrix  (\ref{general}) in the spectral-decomposition  form
\begin{eqnarray}\label{spect}
\rho(t)=\sum_{i=1}^2p_i(t)|w_i(t)\rangle\langle w_i(t)|,
\end{eqnarray}
where
 $p_i(t)$ and $|w_i(t)\rangle$ are the eigenvalues and the eigenvectors of the matrix $\rho(t)$,
respectively. The geometric phase $\Phi(t)$ associated with such an
evolution  is defined as follows  \cite{sjuk2}:
\begin{eqnarray} \label{Phi}
\Phi(t) &=& \arg\left[\sum_{i=1}^2[p_i(0)p_i(t)]^{1/2}\langle
w_i(0)|w_i(t)\rangle\right.\nonumber\\ && \left. \times
\exp(-\int_0^t\langle w_i(s)|\dot{w}_i(s)\rangle ds)\right],
\end{eqnarray}
where  $\arg$  denotes the argument of the complex number,  $\langle
w_i|w_j\rangle$ is a scalar product and   the dot indicates the
derivative with respect to time $s$. For convenience, we assume the
initial time being $t_0=0$. For the sake of completeness
we sketch here, following  Ref. \cite{sjuk2}, the derivation of
Eq.(\ref{Phi}). The  mixed state defined by the density matrix
(\ref{spect}) can be lifted to a pure state $|\Psi(t)\rangle$ in a
larger Hilbert space, i.e.,
\begin{eqnarray}\label{pur}
|\Psi(t)\rangle=\sum_{i=1}^2 \sqrt{p_i(t)}|w_i(t)\rangle\otimes|a_i\rangle,
\end{eqnarray}
where the vectors $|a_i\rangle$ span the Hilbert space of an
arbitrary ancilla. This is known as  a purification of the density
matrix $\rho(t)$ in the sense that $\rho(t)$ is a partial trace of
the density matrix $|\Psi(t)\rangle \langle \Psi(t)|$  over the
ancilla Hilbert space. With the time evolution of the purified
system one can associate the 'Pancharatnam' relative phase
\begin{eqnarray}\label{panch}
\alpha(t)
=  \arg \langle \Psi(0)| \Psi(t)\rangle
\end{eqnarray}
which contains both the gauge--dependent part (a dynamical phase)
and a gauge--independent  part. The central result of \cite{sjuk2}
is to extract from Eq.(\ref{panch}),  by a proper choice of the
'parallel transport condition',   the purification--independent part
which can be termed a geometric phase because  it is  gauge
invariant and reduces to the  known results in the limit of an
unitary evolution \cite{anandan,chruscinski}. The final result is
then given  by  Eq. (\ref{Phi}).

As mentioned in the Introduction, this phase -- contrary to other
attempts of extending the notion of  geometric phase for a
non--unitary evolving quantum system -- has a direct physical
meaning as it can be measured via interferometric experiments
\cite{sjuk2}, i.e. one can construct the purification of the quantum
system such that the relative phase  (\ref{panch}) reduces to the
geometric phase (\ref{Phi}) after suitably defined 'compensating
unitary' cutting of the dynamical part of the relative phase
\cite{sjuk2}.


\section{ Weak coupling regime of  qubit reduced dynamics}

The evolution operator $\Lambda(t,t_0)$  defined by equation
(\ref{general})  or its infinitesimal generator  $\mathcal{L}$
defined by the equation
\begin{eqnarray}\label{gener}
\frac{d}{dt}\rho(t)=\mathcal{L}\{\rho(t)\}
\end{eqnarray}
can be obtained  in a few cases only; namely for stylized, exactly
solvable models or in the limiting regimes such as the  weak
coupling limit or the singular coupling limit \cite{alickilendi}. We
consider a qubit coupled to a bosonic  environment at temperature
$T$. The Hamiltonian of such a system is chosen in the form
\cite{chaos}:
\begin{eqnarray}\label{hamful}
H&=& H_Q\otimes\mathcal{I}+\mathcal{I}\otimes H_B +H_I\otimes V_B, \\
H_B&=&\int_0^\infty E(k)a^\dagger(k)a(k) dk, \\
V_B&=&\int_0^\infty g(k) [a^\dagger(k)+a(k)] dk. \label{inter}
\end{eqnarray}
The operators $ a^\dagger(k)$ and $a(k)$ denote   the creation and
annihilation boson operators, respectively.
 The qubit  Hamiltonian and the interaction  are  assumed to  take the form
\begin{eqnarray}\label{hamQ}
H_Q &=&\frac{\varepsilon}{2}\sigma_z, \,\,\,\,\,\,
H_I=\hbar \mu_x\sigma_x+ \hbar \mu_z\sigma_z,
\end{eqnarray}
where $\sigma_i$ are the Pauli operators, $\varepsilon$ is the qubit
energy splitting and the dimensionless parameters $\mu_x$ and $\mu_z$ are coupling
constants.  Let us remark that if  $\mu_x\equiv 0$ the qubit energy
operator $H_Q$ is an integral of motion,  i.e. it commutes with the
total  Hamiltonian $H$ leaving the expectation value of the
corresponding  energy observable unchanged. This situation defines
the well known exactly solvable  model of pure dephasing \cite{ali}.
A non-vanishing $\mu_x$ is then characteristic for exchange of
energy and related dissipation processes.

For an uncorrelated initial state $\rho(0)\otimes w(\beta)$ taken as
a product of an arbitrary qubit density matrix $\rho(0)$ and the
equilibrium Gibbs state of the environment $w(\beta)= \exp(-\beta
H_B)/\mbox{Tr} [\exp(-\beta H_B)]$ with $\beta=1/k_B T$ ($k_B$ is
the Boltzmann constant), the Davies approximation for the Markovian
kernel yields the following  Markovian master equation
\cite{davies,spohn}
\begin{eqnarray}\label{kin}
\frac{d}{dt}\rho(t)=\mathcal{L}_H\{\rho(t)\}+\mathcal{L}_R\{\rho(t)\},
\end{eqnarray}
where the 'conservative' and  'dissipative' parts read as follows
\begin{eqnarray}\label{kinh}
\mathcal{L}_H\{\rho(t)\}=-\frac{i}{\hbar}[( H_Q+\sum_{k,l=-1}^1 \hbar    s(\Omega_{kl})A^\dagger_{kl}A_{kl} ), \rho(t) ], 
\end{eqnarray}
\begin{eqnarray}\label{kinr}
  \mathcal{L}_R\{\rho(t)\}&=&\frac{1}{2}\sum_{k,l=-1}^1c(\Omega_{kl}) \nonumber \\
&\times &
\left([A_{kl}\rho(t),A_{kl}^\dagger]+[A_{kl},\rho(t)A_{kl}^\dagger]\right),
\end{eqnarray}
where, see in Refs. \cite{alickilendi,lendi},
\begin{eqnarray}\label{kina}
A_{kl}=\mathcal{P}_kH_I\mathcal{P}_l,  \quad
\mathcal{P}_{\pm 1}=|\pm 1\rangle\langle \pm 1|,   \nonumber\\
\Omega_{kl}=  (\lambda_k-\lambda_l) /\hbar, \quad    \lambda_{\pm
1}=\pm \varepsilon/2.
\end{eqnarray}
The states $|1\rangle$ and $|-1\rangle$ denote the excited state and
ground state of the qubit, respectively.  
The quantity $c(\omega)$ is  the Fourier transform of the
autocorrelation function of the bath operator $V_B$ calculated in
the Gibbs state $w(\beta)$ of the bath, namely,
\begin{eqnarray}\label{c}
c(\omega)=\int_{-\infty}^\infty  \mbox{e}^{-i\omega t} \, \mbox{Tr}
[w(\beta)  V_B \mbox{e}^{itH_B/\hbar} V_B \mbox{e}^{-itH_B/\hbar} ]  dt
\end{eqnarray}
and its Hilbert transform  defines the function $s(\omega)$  in the
following way
\begin{eqnarray}\label{s}
s(\omega)=\frac{\mbox{P}}{2\pi}\int_{-\infty}^\infty
\frac{c(x)}{x-\omega}dx,
\end{eqnarray}
where $\mbox{P}$ indicates the Cauchy principal value of the
integral.

In order to treat the complex qubit--environment interaction encoded
in $g(k)$ in  (\ref{inter}) it is convenient  to introduce the
spectral density
\begin{eqnarray}
D(\omega)=\int d k |g(k)|^2   \delta(\omega(k)  -\omega).
\end{eqnarray}
We further limit our consideration to the strictly Ohmic environment
for which  this spectral density is linear with respect to $\omega$
for small frequencies and exhibits an exponential cut-off frequency
$\omega_c$, thereby  exhibiting no non-physical ultraviolet
divergences. Explicitly, this spectral density reads
\begin{eqnarray}\label{J}
D(\omega)=\frac{\alpha}{2}\omega\exp(-\omega/\omega_c),
\end{eqnarray}
where  the dimensionless parameter $\alpha$ characterizes the strength of the environmental
influence on the qubit.
Within this choice    \cite{alickilendi,lendi}
\begin{eqnarray}\label{cs}
c(\omega)=\frac{\pi\alpha}{2} \left(
|\omega|\frac{\exp(\beta\hbar| \omega|)+1}{\exp(\beta \hbar |\omega|)-1}+\omega\right) 
\exp(-|\omega|/\omega_c)
\end{eqnarray}
and $s(\omega)$ is determined via the relation in  (\ref{s}).

 In principle one can solve  Eq. (\ref{kin}) using the
Bloch vector formalism to obtain the coupled evolution equations for
mean values $\langle \sigma_k(t)\rangle, k=x, y, z$  to obtain the
reduced  density matrix as $\rho(t) = (1/2)[1 +\langle
\sigma_x(t)\rangle \sigma_x + \langle \sigma_y(t)\rangle \sigma_y +
\langle \sigma_z(t)\rangle \sigma_z]$. This form allows to extract
the spectral decomposition (\ref{spect}) and the phase $\Phi(t)$.
Such an explicit form of the geometric phase result is, however,
rather cumbersome without exhibiting  much physical insight. We thus
refrain from presenting  such analytical details, but present here
the full analysis of the geometric phase by numerical means. 

%

\section{Analysis of geometric phase}

From Eqs (\ref{general})-(\ref{Phi}) it follows that in order to determine the
geometric phase at arbitrary time $t>0$,  we must specify the
initial state of the qubit. We consider the following  class of initial states
\begin{eqnarray}\label{init1}
|\theta\rangle
=\cos(\theta/2)|1\rangle+\sin(\theta/2)|-1\rangle,
\end{eqnarray}
where $\theta$ is the polar angle in the  Bloch 
sphere  representation. The  corresponding initial
statistical operator $\rho(0)$   takes the  form
\begin{eqnarray}
\rho(0)=|\theta\rangle  \langle\theta|. \label{mat1}
\end{eqnarray}
%
%
\begin{figure}[htpb]
  \begin{center}
\includegraphics[width=0.5\textwidth,angle=270]{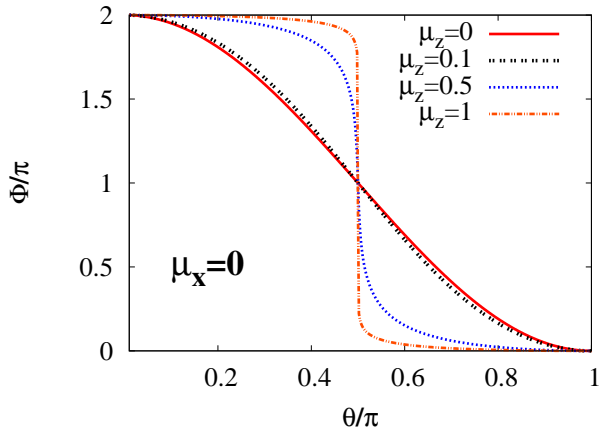}
    \end{center}
  \caption{(Color online) Dependence of the geometric phase $\Phi = \Phi({\cal T})$ on the initial state
of the qubit which is parameterized by the angle $\theta$ on the
Bloch sphere. The qubit is coupled to a purely  dephasing Ohmic
environment, i.e. $\mu_x=0$ in  (\ref{hamQ}). The remaining
parameters are: $\alpha=10^{-2}$,  $\hbar \omega_c/
\varepsilon=10^2$ and $T=0$.  }
  \label{u}
\end{figure}
One of the eigenvalues of this operator is zero, say  $p_2(0)=0$ in
Eq.  (\ref{Phi}), and it  does not contribute to the geometric
phase. This simplifies Eq.  (\ref{Phi}) in that only one term of
the sum survives. The  evolution of the freely evolving qubit, with
$\mu_x=\mu_z=0$ in (\ref{hamQ}), is cyclic with the  time-period
${\cal T}=2\pi \hbar /\varepsilon$ and it acquires the geometric
phase \cite{chruscinski}
\begin{eqnarray}\label{free}
\Phi_0=\pi[1+\cos(\theta)],  \quad   \mbox{mod}  (2\pi),
\end{eqnarray}
which can serve as a  reference for studying the  influence of the
environment.
 In the case of a coupling to an environment, the evolution of the qubit  is not cyclic any longer.
However, below  we consider the phase $\Phi= \Phi({\cal T})$ after
the time ${\cal T}=2\pi \hbar /\varepsilon$ in order to  study the
role of coupling to the environment and for comparison with
(\ref{free}).
\begin{figure}[htpb]
  \begin{center}
 \includegraphics[width=0.5\textwidth,angle=270]{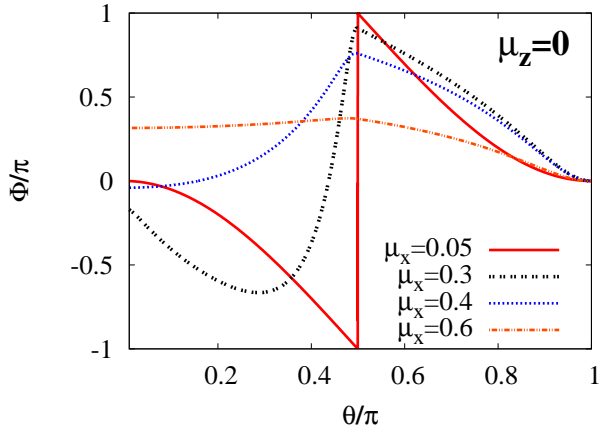}
    \end{center}
  \caption{(Color online)  Geometric phase $\Phi = \Phi({\cal T})$ {\it vs.}
 initial polar angle $\theta$ 
for selected values of the  dissipative qubit - Ohmic environment  coupling strength 
$\mu_x$.  The
dephasing coupling strength  is set at $\mu_z=0$ in (\ref{hamQ}),
and the remaining parameters are the same as in Fig. 1.  }
  \label{w}
\end{figure}

The simplest situation occurs for  pure dephasing; i.e. when
$\mu_x=0$  so that the qubit energy does not change. The results
presented in Fig. \ref{u} show that the geometric phase plotted as a function
of the initial state of the qubit (i.e. as a function of  the
parameter $\theta$ in Eq.  (\ref{init1}))  approaches zero
(modulo $2\pi$) with increasing  coupling  strength $\mu_z$. We
observe that it varies drastically in  the regime near $\theta
=\pi/2$ and varies  weakly outside  this region.  
Moreover, the phase vanishes for $\theta \to 0$  (i.e. for the initially 
excited state $|1\rangle$ ) and $\theta \to \pi$  (i.e. for the ground
state $|-1\rangle$ ). This finding corroborates the  results for the phase in the
exactly solvable  model  of   pure dephasing with arbitrary (not only weak)  coupling  \cite{myfaz}.  
For the presentation as in Fig. 1, we note that  the function $\Phi \left(\theta\right)$  is antisymmetric about the point $\{\theta, \Phi\} = \{\pi/2, \pi\}$, i.e. the relation  
\begin{eqnarray}\label{symm}
\Phi \left(\frac{\pi}{2} + \theta\right) = 2\pi -  \Phi \left(\frac{\pi}{2} -  \theta\right) \quad 
\mbox{for} \quad \theta \in[0, \pi/2]
\end{eqnarray}
holds.  It can be interpreted as a rotation  symmetry around the point $\{\theta,\Phi\}=\{\pi/2,\pi\}$. 

\begin{figure}[htpb]
  \begin{center}
  \includegraphics[width=0.5\textwidth,angle=270]{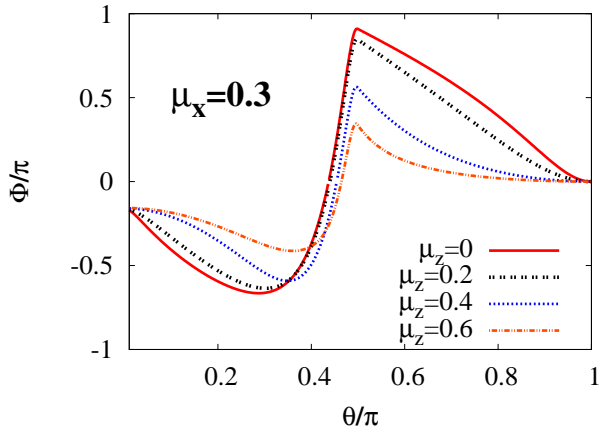}
    \end{center}
  \caption{(Color online) Role of a qubit-environment coupling on the  geometric phase  $\Phi = \Phi({\cal T})$
{\it vs.} initial state preparation   $\theta$ for a qubit that is
coupled to an Ohmic environment. The dissipative coupling strength
is set at $\mu_x=0.3$. The influence of dephasing is depicted for four
different coupling strengths $\mu_z$. 
The remaining parameters are the same as in Fig. 1. }
  \label{wu}
\end{figure}

A  most intriguing behavior on the role of the environment
emerges when $\mu_x \neq 0$; i.e. when  the qubit--environment interaction  is
allowed to exchange energy with the qubit system.  The results
depicted  in Fig. \ref{w} show the qualitative changes in the
geometric phase properties for increasing  dissipation coupling strength $\mu_x$ as a function of the polar angle $\theta$.
Note that the geometric phase $\Phi$  in Fig.  \ref{w}  is plotted differently
 from Fig.  \ref{u} with $\Phi/\pi$ varying within the interval $ (-1,1)$.  We have decided to make this change  in order to avoid confusing jump-like behavior of $\Phi(\theta)$  in vicinity of the polar angle $\theta=\pi/2$.  E. g.  the curve corresponding to the case $\mu_x=0.05$ in Fig.   \ref{w}  is very similar to the curve corresponding to the case $\mu_z=0.5$ in Fig.   \ref{u}. However,  presented in the  
$\Phi/\pi\in (-1,1)$  interval it exhibits jump-like  behavior which is an artefact of the way the plot is done.  Let us recall again here that the phase $\Phi$ is defined modulo $2\pi$. 

For  small values of $\mu_x$, the geometric
phase is close to that for the isolated qubit, cf.  $\mu_x =0.05$ in
Fig. \ref{w}  when compared with Fig. \ref{u}  but with $\Phi$ varying there
$\Phi/\pi \in (0, 2)$.  When $\mu_x$ increases the function $\Phi(\theta)$  exhibits a local maximum and minimum, see the case $\mu_x =0.3$  in Fig. \ref{w}.  For larger value of the coupling strength $\mu_x$  (the case  
$\mu_x =0.4$ )  the geometric phase is an increasing function of the polar angle $\theta$ till to the value $\theta=\pi/2$  reaching a local maximum. Next, it decreases as  $\theta \to \pi$. 
In comparison to the dephasing coupling, in this case we can find at least three distinguishing features of the geometric phase.  
Firstly, we note  breaking of antisymmetry  of $\Phi(\theta)$,   being in
distinct contrast to the case of pure dephasing ($\mu_x = 0$), cf.  Fig. 1. Secondly, 
 the dependence of
the phase on the initial state parameterized by $\theta$ is
non-monotonic, exhibiting a local maximum and a minimum.
Thirdly, the geometric phase $\Phi$ vanishes for $\theta
\to \pi$ (i.e. for the ground state) but not necessary so for
$\theta \to 0$ (i.e. for the excited state). 

One can observe that for a fixed $\mu_x$,  the dephasing process controlled by $\mu_z$ does
 not change the qualitative properties of the geometric phase $\Phi$, see Figs. 1 and 3.
Pure dephasing affects only the off--diagonal elements of the
density matrix,  becoming closer to the maximally mixed qubit state.
The geometric phase in a quantum evolution of such  states vanishes.
In a general energy--exchanging process the time dependence of the
density matrix is more complex and the geometric phase $\Phi$
seemingly quantifies this fact. 
Moreover, the stability of geometric phase with respect to
decoherence is crucial for effectiveness of holonomic quantum
computation \cite{zanardi,yellow}. It is evident that the stability
of phase can be significantly improved via a proper choice of  the
initial state determined  by $\theta$ in Eq. (\ref{init1}).
\begin{figure}[htpb]
  \begin{center}
     \includegraphics[width=0.5\textwidth,angle=270]{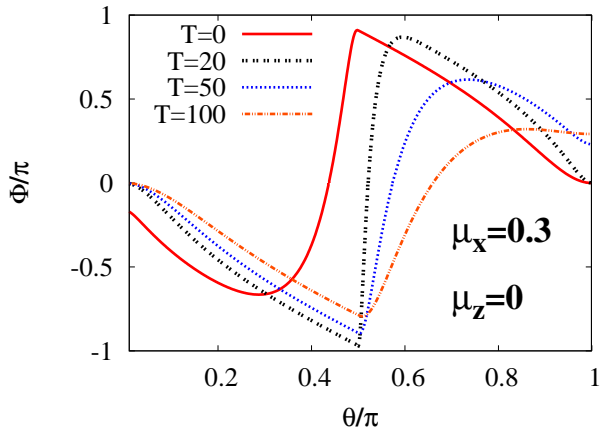}
    \end{center}
  \caption{(Color online) The influence of varying
temperature $T$ on the geometric phase $\Phi({\cal T})$ {\it vs.}
initial state  preparation $\theta$  is depicted for a dissipative
qubit with $\mu_x=0.3$  and zero
dephasing, i.e. $\mu_z=0$. The remaining parameters are   the same
as in Fig. 1.  Temperature is measured in units of
$\varepsilon/k_B$.} \label{temp}
\end{figure}

Thus far we considered  zero temperature, $T=0$.  The effect of
increasing temperature is depicted in Fig. \ref{temp}. Firstly, we
observe  that if temperature increases the phase does not vanish for $\theta \to \pi$ while it
tends to zero  for $\theta \to 0$.   Secondly, the main properties remain similar: In all
presented cases  a minimum and a maximum exist. However, the maximum  
diminishes with  increasing temperature. 


\section{Concluding remarks}

No realistic physical quantum system is in perfect isolation from
its environment. At best one can achieve a weak coupling between the
system and the environment. In this  weak coupling regime it is
possible to extract the reduced dynamics of the open quantum system
in a mathematically satisfactory and controlled  way by using a
Markovian reduced dynamics following the Davies scheme. In this work
we have analyzed the geometric phase of a qubit in the presence of a
weak coupling to a bosonic environment.
We have investigated the relation between the geometric phase $\Phi$
and the mechanism for decoherence of the qubit for either the case
of pure dephasing with $\mu_x = 0$ or in presence of dissipative
energy relaxation, i.e. $\mu_x \neq 0$. The latter situation allows
for a significant variation of the emerging geometric phase upon
varying the coupling strength $\mu_x$. 
A variation of  the  pure dephasing coupling, i.e. $\mu_z\neq 0$
with $\mu_x=0$, between qubit and   environment  barely affects the
geometric phase. This feature is distinct from other set-ups, such
as the emergence of  quantum entanglement in open systems, where
this dephasing-coupling mechanism  can play a dominant or a similar
role as an energy relaxation-coupling.

Nowadays, the geometric  phase plays a crucial role in a variety of physical problems and has observable consequences in a wide range of  systems. 
Under various aspects, this concept occurs in geometry, astronomy, classical mechanics, and quantum theory. 
 The impressive recent  progress in nanotechnology and
experimental techniques allows one to test the fundamentals of
quantum dynamics and details of interactions modeled by
Hamiltonians.    The geometric phase is not a quantum mechanical
observable, i.e. it is not represented by a Hermitian operator.
However, it can be experimentally measured, cf. Ref. \cite{exper}. 
It can be used to encode information on systems. E.g. it has been proposed as an order parameter for quantum phase transitions \cite{ochnik}.  
The results obtained  here  suggest that one can  also exploit the
geometric phase as a  quantifier characterizing  a nature of the
system-environment coupling. Indeed, three features of the geometric
phase $\Phi= \Phi({\cal T})$ allow one to distinguish the character
of qubit-environment coupling (i.e. pure dephasing {\it vs.}
dissipation): (i) rotation symmetry around some  points on the 
$\theta-\Phi$ plane or equivalently antisymmetric dependence of $\Phi$  upon  $\theta$ about some points, (ii) non-monotonic behavior of $\Phi$
with respect to $\theta$ and (iii) the behavior of $\Phi$ for $\theta \to 0$ (i.e.
 for the qubit prepared in the  excited state). We have verified that all these three features
are manifest also  at  times $t=n {\cal T} (n=2, 3, 4)$  for the measurable quantifier $\Phi= \Phi({n \cal T})$. 
This feature of the geometric phase $\Phi$ thus presents an
additional suitable tool in exploring characteristics of open system
interactions at the quantum scale.

\section*{Acknowledgment} Work  supported by the Polish Ministry of Science
and Higher Education under the grant N 202 131 32/3786, the German
Excellence Initiative via the ``Nanosystems Initiative Munich
(NIM)'' and by the DFG-SFB 631.



\end{document}